\documentclass[sigconf]{acmart}

\usepackage{enumitem}

\usepackage{cleveref}
\crefname{figure}{Figure}{Figures}

\AtBeginDocument{%
  }

\copyrightyear{2026}
\acmYear{2026}
\setcopyright{cc}
\setcctype{by}
\acmConference[WWW '26] {Proceedings of the ACM Web Conference 2026}{April 13--17, 2026}{Dubai, United Arab Emirates.}
\acmBooktitle{Proceedings of the ACM Web Conference 2026 (WWW '26), April 13--17, 2026, Dubai, United Arab Emirates}
\acmISBN{979-8-4007-2307-0/2026/04}
\acmDOI{10.1145/3774904.3792801}

\newcommand{\zyx}[1]{\textcolor[rgb]{0.2, 0.8, 0.2}{\textbf{#1}}}
\newcommand{\xrz}[1]{\textcolor{red}{#1}}

\renewcommand{\zyx}[1]{#1}
\renewcommand{\xrz}[1]{#1}

\begin{document}

\title{From Reasoning LLMs to BERT: A Two-Stage Distillation Framework for Search Relevance}

\author{Runze Xia}
\authornote{Equal contribution.}

\affiliation{%
    \institution{Meituan}
  \city{Shanghai}
  \country{China}
}
\email{runze019@gmail.com}

\author{Yupeng Ji}
\authornotemark[1]
\affiliation{%
  \institution{Meituan}
  \city{Shanghai}
  \country{China}
}
\email{jiyupeng02@meituan.com}

\author{Yuxi Zhou}
\authornotemark[1]
\affiliation{%
  \institution{Meituan}
  \city{Shanghai}
  \country{China}
}
\email{zhouyuxi05@meituan.com}

\author{Haodong Liu}
\affiliation{%
  \institution{Meituan}
  \city{Shanghai}
  \country{China}
}
\email{liuhaodong05@meituan.com}

\author{Teng Zhang}

\affiliation{%
  \institution{Meituan}
  \city{Shanghai}
  \country{China}
}
\email{zhangteng09@meituan.com}

\author{Piji Li}
\authornote{Corresponding author.} 
\affiliation{%
  \institution{Researcher}
  \city{Nanjing}
  \country{China}
}
\email{lipiji.pz@gmail.com}

\begin{abstract}

Query-service relevance prediction in e-commerce search systems faces strict latency requirements that prevent the direct application of Large Language Models (LLMs). To bridge this gap, we propose a two-stage reasoning distillation framework to transfer reasoning capabilities from a powerful teacher LLM to a lightweight, deployment-friendly student model. In the first stage, we address the limitations of general-purpose LLMs by constructing a domain-adapted teacher model. This is achieved through a three-step process: domain-adaptive pre-training to inject platform knowledge, supervised fine-tuning to elicit reasoning skills, and preference optimization with a multi-dimensional reward model to ensure the generation of reliable and preference-aligned reasoning paths. This teacher can then automatically annotate massive query-service pairs from search logs with both relevance labels and reasoning chains. In the second stage, to address the challenges of architectural heterogeneity in standard distillation, we introduce Contrastive Reasoning Self-Distillation (CRSD). By modeling the behavior of the same student model under ``standard'' and ``reasoning-augmented'' inputs as a teacher-student relationship, CRSD enables the lightweight model to internalize the teacher's complex decision-making mechanisms without needing the explicit reasoning path at inference. Offline evaluations and online A/B testing in the Meituan search advertising system demonstrate that our framework achieves significant improvements across multiple metrics, validating its effectiveness and practical value.
\end{abstract}



\maketitle

\section{Introduction}
In the realm of e-commerce, the core mission of a search engine is to precisely connect user queries with a vast repository of products, making query-service relevance prediction a cornerstone technology for achieving this goal. In large-scale e-commerce search systems, the search process typically employs a multi-stage architecture \cite{guo2022semantic, momma2022multi}: A retrieval module (coarse ranking) selects a candidate set from the massive item repository; subsequently, a relevance model within the fine-grained ranking stage conducts a detailed evaluation of each <query, service> pair, filtering out irrelevant results to ensure quality and user experience of the final-displayed products. As the relevance model must handle hundreds of millions of query-service evaluations daily, it must possess exceptional inference efficiency while maintaining high accuracy \cite{khattab2020colbert}.

Traditional relevance models, such as those based on BERT's encoder-only architecture \cite{devlin2019bert}, excel at capturing surface-level semantic connections between queries and services through pre-trained language representations. However,
{despite their high inference efficiency, }
these discriminative models generally lack deep reasoning capabilities, rendering them ill-equipped for complex semantic understanding and logical judgment tasks. For instance, when a user searches for “skin rejuvenation treatment” and a merchant offers “facial hydrotherapy session,” the two phrases appear entirely different in wording yet describe essentially the same type of beauty service. Traditional models often fail to recognize such semantic equivalence grounded in real-world knowledge, leading to the erroneous exclusion of highly relevant results.

In recent years, the remarkable capabilities of Large Language Models (LLMs) have offered a promising new direction for tackling these challenges \cite{wei2022chain,guo2025deepseek,zhao2023survey}. Through large-scale pre-training, these models internalize extensive world knowledge, serving as the foundation for their advanced abilities in natural language understanding and logical reasoning \cite{xu2025decider}. This foundation enables them not only predict relevance labels, facilitating the large-scale generation of training data to substitute for manual annotation, but also provide detailed reasoning paths to explain their judgments upon request. This ``reasoning process'' or ``chain-of-thought'' represents the cognitive steps the model takes to reach a specific decision. Such a reasoning ability provides interpretability for relevance prediction and has the potential to yield more accurate and trustworthy outcomes.

However, the direct application of LLMs in industrial-grade search systems faces significant challenges. First, latency and cost: The enormous model size and computational demands of LLMs make it infeasible to meet the strict millisecond-level response times required in production environments. Second, domain knowledge deficiency: General-purpose LLMs lack an understanding of platform-specific products, scenarios, and business rules, often leading to suboptimal performance. Third, inconsistent standards: Different platforms have unique definitions and preferences for relevance, making the direct application of a generic LLM ineffective.

Furthermore, while lightweight models like BERT can satisfy latency requirements and are widely used in industry, their heavy reliance on data quality and inherent deficiencies in general knowledge and reasoning capabilities are significant limitations. The prevailing approach to bridge this gap is knowledge distillation—
using a teacher model to transfer its knowledge to a small model. However, traditional distillation methods face two key technical obstacles when transferring reasoning abilities: (1) Architectural Heterogeneity: The disparity between decoder-only LLMs and encoder-only BERT models makes direct feature alignment difficult \cite{hofstatter2020improving}. 
(2) Focus on Output Imitation over Process Understanding: Existing methods primarily focus on mimicking the final output, failing to effectively transfer the intrinsic logical mechanisms of the reasoning process.

To address these critical issues, this paper proposes a two-stage reasoning distillation framework designed to efficiently transfer the reasoning capabilities of a large model to a deployment-friendly lightweight model while preserving computational efficiency.

Stage One: Building a Domain-Adapted Reasoning LLM. To address the shortcomings of general-purpose LLMs in platform-specific scenarios, we construct a specialized domain reasoning LLM. This is achieved through a \textbf{three-step training pipeline}: continuous pre-training to inject domain knowledge, supervised fine-tuning to elicit reasoning abilities, and preference optimization using a reward model for fine-grained enhancement of reasoning quality. This model not only overcomes the limitations of generic models but also generates high-quality relevance labels and detailed reasoning paths for massive query-service pairs.

Stage Two: Contrastive Reasoning Self-Distillation. We propose the Contrastive Reasoning Self-Distillation (CRSD) method, which models the behavior of the same BERT model under standard input and reasoning-augmented input configurations as a teacher-student relationship. Through the InfoNCE contrastive learning mechanism \cite{oord2018representation, gao2021simcse}, this method effectively resolves the challenges of feature alignment between heterogeneous architectures. It enables the lightweight model to internalize the reasoning capability without needing the explicit reasoning path during inference \cite{zhang2019your}.

We conducted rigorous evaluations on a large-scale, real-world dataset and validated the method's effectiveness through online A/B testing within Meituan's e-commerce search system. The results demonstrate that our approach significantly improves relevance prediction accuracy while maintaining inference efficiency, offering a practical solution for reasoning capability transfer in industrial applications. Our main contributions are threefold:

\begin{itemize}[leftmargin=2em,topsep=0pt,partopsep=0pt]
\item \textbf{A Domain-Adapted Reasoning LLM Pipeline:} We propose a three-step pipeline to construct a reasoning LLM tailored for e-commerce, enabling it to generate high-quality relevance judgments and detailed reasoning paths.
\item \textbf{A Novel Contrastive Reasoning Distillation Framework (CRSD):} We introduce CRSD, a framework that effectively transfers LLM reasoning to a lightweight model by overcoming the critical challenges of architectural heterogeneity.
\item \textbf{Comprehensive Validation and Production Deployment:} We demonstrate the effectiveness of our approach through extensive experiments on real-world datasets and a successful online deployment in a large-scale e-commerce search engine.

\end{itemize}

\section{Related Work}
\begin{figure*}[t!]
    \centering
    \includegraphics[width=\textwidth]{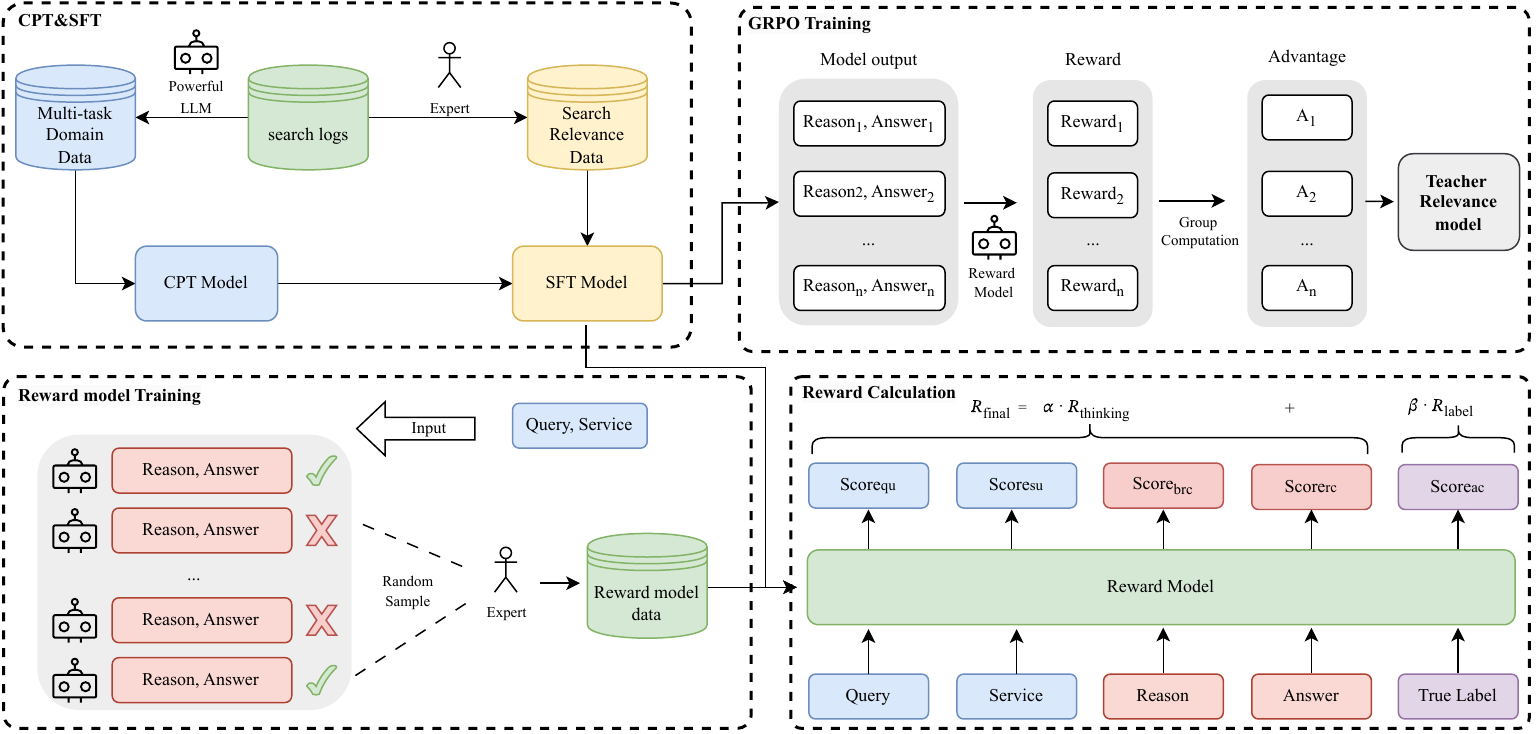}
    \caption{An overview of our three-stage training pipeline for the domain-adapted reasoning LLM, consisting of Continued Pre-training (CPT), Supervised Fine-tuning (SFT), and preference optimization.}
    \label{fig:llm}

\end{figure*}

\subsection{Search Relevance in E-commerce}
Search relevance is a fundamental problem in e-commerce platforms, directly influencing user experience and business revenue. The objective is to determine the degree of semantic alignment between a user query and a candidate service (e.g., product or service description). In practical systems, relevance is not a binary concept but a graded one. To better capture this real-world ambiguity, we adopt a three-level classification scheme consisting of irrelevant, moderate, and relevant categories. This formulation reflects the nuanced relationship between user intent and item semantics. 

Traditional retrieval models, such as the probabilistic relevance framework (BM25) \cite{robertson2009probabilistic}, have long been used to measure query-service similarity. However, these approaches rely heavily on lexical matching and often fail to capture deeper semantic relevance. To address this limitation, deep learning–based methods have been proposed. Huang et al. \cite{huang2013learning} introduced the Deep Structured Semantic Model (DSSM), which learns distributed representations of queries and services from click-through data. With the rise of pre-trained language models, Transformer-based architectures have shown remarkable performance improvements. Nogueira and Cho \cite{nogueira2019passage} demonstrated that fine-tuning BERT for passage re-ranking substantially enhances ranking accuracy, validating the effectiveness of contextualized embeddings in relevance modeling.

Building on these advances, large-scale e-commerce platforms have incorporated neural relevance models into their production search engines. For example, Walmart has deployed semantic retrieval and entity-aware multi-task models to improve query understanding and ranking \cite{magnani2022semantic,peng2023entity}, while Amazon has enhanced product retrieval through semantic product search and query understanding frameworks \cite{nigam2019semantic,luo2024exploring}. Similarly, Taobao introduced large language model–based query rewriting for long-tail search \cite{peng2024large}, and JD.com optimized multi-stage, multi-grained semantic embeddings for product retrieval \cite{wang2023learning}. These industrial deployments demonstrate the practical value and scalability of semantic relevance modeling in real-world e-commerce search systems.

\subsection{Reasoning Models for Search Relevance}
The emergence of large language models (LLMs) has significantly advanced reasoning and semantic understanding in search relevance tasks. Models such as GPT-4~\cite{achiam2023gpt} and Qwen~\cite{yang2025qwen3} exhibit strong contextual understanding and few-shot generalization capabilities, enabling more accurate alignment between user intent and service semantics. Recent reasoning-enhanced architectures, such as chain-of-thought (CoT) prompting~\cite{wei2022chain} and DeepSeek-R1~\cite{guo2025deepseek}, explicitly model multi-step logical inference, offering a pathway to more interpretable and robust relevance estimation.

Building upon these reasoning capabilitaies, several industrial systems have begun integrating LLMs for search relevance prediction. Wang et al.~\cite{wang2024improving} show improvements in user preference modeling at Pinterest. However, the high latency and cost of LLM inference limit their real-time deployment in production systems.

\subsection{Knowledge Distillation for Search Relevance} \label{sec:kd}
Knowledge distillation (KD) provides an efficient framework for transferring knowledge from large teacher models to compact students. Hinton et al. \cite{hinton2015distilling} introduced this paradigm, and subsequent models such as DistilBERT \cite{sanh2019distilbert}, TinyBERT \cite{jiao2019tinybert}, and MobileBERT \cite{sun2020mobilebert} showed that Transformer-based students can retain strong accuracy while greatly improving inference efficiency. 

In information retrieval and e-commerce search, knowledge distillation has been widely adopted to compress heavy cross-encoders into lightweight bi-encoder or dual-tower models\zyx{~\cite{hofstatter2020improving,yao2022reprbert,zhang2022distilling,liu2022knowledge}}. 
Building on this foundation, several large-scale e-commerce platforms have integrated KD frameworks into production search systems. For instance, ReprBERT~\cite{yao2022reprbert} was successfully applied to Alibaba’s product retrieval pipeline, and Liu et al.~\cite{liu2022knowledge} designed a contextual distillation model to enhance relevance matching in e-commerce product search.

Shang et al. \cite{shang2025knowledge} leveraged LLM-based distillation to enhance Walmart’s e-commerce search, demonstrating that student models distilled from reasoning-rich teachers can achieve strong performance under strict latency constraints. Extending this direction, Ye et al. \cite{ye2024best} proposed DisRanker, which integrates point-wise and margin-MSE objectives for web-scale ranking distillation. Zhao et al. \cite{zhao2025explainable} introduced a multi-dimensional framework that jointly transfers score distributions and chain-of-thought representations to improve interpretability and relevance consistency. Closely related to our work, Agrawal et al. \cite{agrawal2025rationale} encoded reasoning paths generated by an external LLM using a separate sentence encoder and aligned them with BERT’s \texttt{[CLS]} vector. However, this approach overlooks that the reasoning LLM may not align well with domain-specific data, leading to suboptimal label accuracy.

\section{The Reasoning then Distilling Framework}
\subsection{Framework Overview}

In Meituan's search relevance prediction task, given a user query and a candidate service, the objective is formulated as predicting the relevance degree between them, which is typically a multi-class label (e.g., 0-irrelevant, 1-moderate, 2-relevant).

We proposed \textbf{Reasoning-then-Distilling framework} in this paper, which operates in two primary stages. The first stage is dedicated to building a powerful teacher model $M_t$. As can be seen in \Cref{fig:llm}, this is achieved through domain-adaptive continued pre-training, supervised fine-tuning, and preference optimization guided by a multi-dimensional reward model, collectively ensuring that the teacher model delivers not only accurate relevance labels but also trustworthy reasoning chain. Subsequently, the second stage involves training a lightweight student model $M_s$ via a contrastive reasoning self-distillation approach, enabling it to internalize the teacher's complex decision-making mechanisms without explicitly generating intermediate reasoning steps.

\subsection{Search Relevance Reasoning LLM Training Stage}
\subsubsection{Domain-Adaptive Continued Pre-Training (CPT)} 

To enhance an open-source LLM with domain-specific knowledge (e.g., platform-specific service information), we perform domain-adaptive continued pre-training. For this purpose, we first construct a large-scale, multi-task search relevance corpus. The construction of this corpus involves sourcing data from online search logs and then employing several powerful LLMs for automatic annotation.  The resulting corpus comprises millions of samples across seven task types, as detailed in ~\Cref{tab:pretrain task}. Subsequently, we performed continued pre-training on an open-source base LLM using this multi-task corpus, which yields the model $M_{cpt}$. This model is now equipped with the foundational domain knowledge for complex reasoning tasks.

\begin{table}[h]
\centering
\caption{Overview of Task Types and Descriptions in the Continue Pre-training Corpus}

\begin{tabular}{|l|p{0.6\linewidth}|}
\hline
\textbf{Task} & \textbf{Description} \\
\hline
Query Understanding & Given a user query, please describe the user intent. \\
\hline
Service Understanding & Given a platform service, please describe the service. \\
\hline
Query2service & Given a user query, please generate a service that may be clicked. \\
\hline
Service2query & Given a platform service, please generate a query that may be clicked. \\
\hline
Pointwise Relevance & Given a user query and a platform service, please predict the relevance degree. \\
\hline
Pairwise Relevance & Given a user query and a pair of  platform services , please predict which one is more relevant.\\
\hline
Setwise Relevance & Given a user query and a list of  platform services , please predict which one is most relevant. \\
\hline
\end{tabular}

\label{tab:pretrain task}

\end{table}

\subsubsection{Supervised Fine-Tuning (SFT)}

The goal of this stage is to align the base model with human-like reasoning. To this end, we first curated a high-quality dataset of tens of thousands of samples, each comprising a <query, service> pair and a meticulously human-annotated reasoning chain leading to a final answer.

We then fine-tuned the base model $M_{cpt}$  on this dataset, enabling it to predict the corresponding reasoning chain $r_i$ and answer $a_i$ jointly from the input pair $(q_i, s_i)$, where $q_i$ and $s_i$ represent the $i$-th \textbf{query} and \textbf{service}, respectively. The model is optimized by minimizing the standard cross-entropy loss over the sequence:
\begin{equation}
    \mathcal{L}_{\text{SFT}} = -\sum_{i=1}^{N} \left[ \log P_{\theta}(r_i, a_i \mid q_i, s_i) \right]
\end{equation}
where $N$ is the total number of training samples. For the remainder of this paper, we consistently use $q$ and $s$ to denote query and service.

This alignment process compels the model to internalize the human-annotated reasoning patterns. 
Consequently, this model, denoted as $M_{SFT}$, gains the preliminary ability to produce a logical and business-rule-compliant reasoning chain before delivering its final answer.

\subsubsection{Preference Optimization with a Multi-Dimensional Reward Model} 

While supervised fine-tuning instills human-like reasoning patterns, it does not guarantee logical quality or reliability. 
To address this limitation, we employ reinforcement learning (RL) to further optimize the model's reasoning capabilities.  However, a critical challenge in applying RL is the design of an effective reward function\zyx{~\cite{ng1999policy, randlov1998learning, tu2025informative}}. 
A naive approach, such as rewarding only the final answer's correctness, suffers from two critical issues. \textbf{Firstly}, the reward signal is exceedingly sparse, providing insufficient guidance for the multi-step reasoning process~\cite{zhang2025100}. \textbf{Secondly}, it is highly susceptible to ``reward hacking,'' where the model might generate correct answers through entirely flawed or coincidental reasoning chains, thus failing to achieve the true goal of reliable inference~\cite{tarek2025reward}.

To address this, we proposed a fine-grained reward model$M_{\text{RM}}$ to simultaneously evaluates both the reasoning process and the final outcome, providing richer and denser guidance signals. This reward model was then used to optimize our policy model($M_{SFT}$) via the GRPO algorithm~\cite{guo2025deepseek}, thereby enhancing its reasoning capabilities.

\textbf{Reward Model Construction}. Training a model capable of precisely assessing reasoning quality requires a high-quality, expert-annotated dataset. We first construct this by using multiple LLMs of varying scales to generate diverse <reason, answer> pairs for a given <query, service> input. From this pairs, we sample a wide spectrum of reasoning styles and potential error modes. Subsequently, domain experts meticulously annotate each output along the following five dimensions (scored 0-4): 1) \textbf{Query Understanding}: Accuracy in capturing the user's search intent.
2) \textbf{Service Understanding}: Accuracy in describing the core attributes of the service.
3) \textbf{Business Rule Compliance}: Adherence to the platform's business logic during reasoning.
4) \textbf{Reasoning Consistency}: The degree to which the reasoning rationale logically supports the final answer.
5) \textbf{Answer Correctness}: Whether the generated answer matches the ground-truth label.

The reward model $M_{RM}$, initialized from the SFT model $M_{SFT}$, is then trained to predict these five expert-annotated scores via supervised fine-tuning. Specifically, it takes  <query, service, reason, answer with true\_label> as input and is fine-tuned to output the corresponding five-dimensional score vector. On a held-out test set, $M_{RM}$ achieves $89.77\%$  agreement rate with expert evaluations,  confirming its reliability as a reward function.

\textbf{GRPO with Weighted Reward}. To refine the reasoning capabilities of the SFT model, we introduce a reinforcement learning stage that leverages the GRPO algorithm. A central component of this stage is our proposed multi-dimensional reward model, which provides a structured learning signal by disentangling the evaluation of the reasoning process from that of the final outcome. Formally, For a given prompt $x=(q,s)$, the policy model samples $N$ independent outputs ${(r_i, a_i)\}_{i=1}^{N}}$. Each output is scored by $M_{RM}$, after which we compute a process reward $R_{thinking}$ (the average of the first four dimension scores) and a label reward $R_{label}$  (the fifth dimension score). The final composite reward is calculated as:
\begin{equation}
    R_{\text{final}} = \alpha \cdot R_{\text{thinking}} + \beta \cdot R_{\text{label}}
    \label{eq:final_reward}
\end{equation}

The GRPO objective function is defined as:

\begin{equation}
\small
\label{eq:grpoloss_compact}
\begin{aligned}
\mathcal{J}_\text{GRPO}(\theta) = \mathbb{E}_{\substack{(q,a)\sim \mathcal{D} \\ \{o_i\}_{i=1}^G\sim \pi_{\theta_\text{old}}(\cdot\mid q)}} \Bigg[ & \frac{1}{G}\sum_{i=1}^{G} \frac{1}{|o_i|}\sum_{t=1}^{|o_i|} \bigg( \\
& \min \Big( \rho_{i,t}(\theta) \hat{A}_{i,t}, \\
& \qquad \text{clip} \big( \rho_{i,t}(\theta), 1 - \varepsilon, 1 + \varepsilon \big) \hat{A}_{i,t} \Big) \\
& - \beta D_{\text{KL}}(\pi_{\theta} || \pi_{\text{ref}}) \bigg) \Bigg],
\end{aligned}
\end{equation}
where
\begin{equation}
    \rho_{i,t}(\theta)=\frac{\pi_{\theta}(o_{i,t} \mid q, o_{i,<t})}{\pi_{\theta_{\text{old}}}(o_{i,t} \mid q,o_{i,<t})}.
\end{equation}

Here, the advantage function 

$\hat{A}_{i,t} = \frac{R_{\text{final}}^i - \text{mean}(\{R_{\text{final}}^i\}_{i=1}^G)}{\text{std}(\{R_{\text{final}}^i\}_{i=1}^G)}$
measures the relative quality of the \(i\)-th output within its group $G$.

This multi-dimensional weighted reward design ensures that for different reasoning paths of the same problem, the model receives optimization signals that are denser and more fine-grained than those from a single outcome reward. It explicitly guides the policy model not merely to pursue correct answers, but to converge steadily towards a reasoning pattern characterized by a reliable process and accurate results.
\subsection{Reasoning Distillation Stage}

The domain-adapted reasoning LLM described previously can generate high-quality labels and reasoning paths for a massive volume of <query, service> pairs from platform search logs, providing abundant data for training lightweight online models. To meet the low-latency requirements of our production environment, we select a 6-layer BERT as the target model for online deployment. To fully leverage the valuable reasoning paths generated by the LLM, we design a novel reasoning distillation framework to efficiently transfer the LLM's reasoning capabilities to this lightweight BERT model.

\subsubsection{Baseline Reasoning Distillation Method} \label{sec:baseline_distill}

Distilling knowledge from an LLM to a lightweight BERT model faces two fundamental challenges. Architectural Heterogeneity: The vast differences in architecture and representation space between a decoder-only LLM and an encoder-only BERT make direct feature-layer alignment exceedingly difficult. 

\begin{figure}[h!]
    \centering
    \includegraphics[width=0.9\columnwidth]{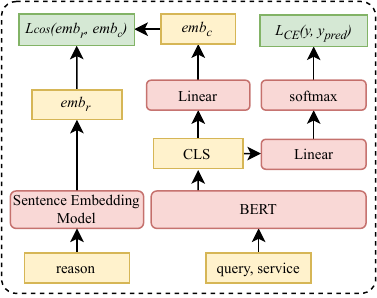}
    \caption{An illustration of the baseline reasoning distillation method.}
    \label{fig:baseline_distill}
    \vspace{-1em}
\end{figure}

Given these challenges, a common approach is to use the reasoning path as an auxiliary training signal through a dual-objective framework, as illustrated in ~\Cref{fig:baseline_distill}. In this method, a pre-trained sentence embedding model first encodes the reasoning path $r$ into a dense vector, $\mathbf{emb}_r$. Concurrently, the BERT model processes the standard \texttt{<query, service>} pair to produce the \texttt{[CLS]} representation, $\mathbf{cls}$. This $\mathbf{cls}$ vector is then utilized in two parallel tasks. For the primary classification task, it is passed through a standard classification head (a linear layer and a softmax function) to produce the prediction $y_{pred}$, which is trained with a cross-entropy loss, $\mathcal{L}_{CE}(Y, y_{pred})$. Simultaneously, for the reasoning alignment task, the $\mathbf{cls}$ vector is projected by another linear layer into a new vector, $\mathbf{emb}_c$. This vector is then encouraged to align with the reasoning vector $\mathbf{emb}_r$ using a cosine similarity-based loss, denoted as $\mathcal{L}_{\text{cos}}$. The final training objective is a weighted sum of these two losses, formulated as:
\begin{equation}
    \mathcal{L}_{\text{total}} = \mathcal{L}_{CE} + \mu \mathcal{L}_{\text{cos}}
\end{equation}
where $\mu$ is a balancing hyperparameter.

Although this approach is intuitive, it suffers from a fundamental limitation highlighted in Section \ref{sec:kd}: the representational gap between heterogeneous models. Because it attempts to align representations from two independently trained models with misaligned semantic spaces (an external sentence encoder and the BERT model), the cross-model supervision signal is indirect. This makes the knowledge transfer more akin to surface-level imitation rather than a deep internalization of the logical process. Consequently, the model struggles to truly integrate the reasoning flow into its own parameters—the core challenge our self-distillation approach is designed to overcome.

\begin{table}[h!]
\centering
\caption{An example illustrating the nuanced relevance judgment required in e-commerce search.}
\vspace{-3mm}
\begin{tabular}{|l|p{0.75\linewidth}|}
\hline
\textbf{Query} & Freshly-made cocktail \\
\hline
\textbf{Service} & \textbf{Name:} Jiu Xiang Homebar  

\textbf{Tags:} cocktail set, single cocktail, whiskey coke bucket, all-you-can-drink, ... \\
\hline
\textbf{Reasoning} & The user's query ``Freshly-made cocktail'' implies an intent for on-site preparation. The service, a 'Homebar', not only explicitly mentions 'cocktail set' but also belongs to the 'Bar' category. The key inference is that cocktails in a bar context are inherently made-to-order, thus satisfying the 'freshly-made' requirement... Therefore, the service is highly relevant. \\
\hline
\textbf{Label} & 2 \\
\hline
\end{tabular}
\label{tab:example}
\end{table}

\subsubsection{Contrastive Reasoning Self-Distillation}

Our primary goal is to enable the lightweight model to \textit{internalize} and apply the LLM's logical judgment capabilities during inference. We posit that the key logic and decisive factors embedded in the reasoning path $r$ are central to enhancing the model's relevance prediction ability. However, in an online setting, the model only has access to the standard \texttt{<query, service>} input.

Based on this, we propose a core hypothesis: if a model learns to process a reasoning-augmented input \texttt{<query, service, reason>} during training, this acquired reasoning pattern can be \textit{internalized} into the model's parameters through a suitable mechanism. Ultimately, even when given only the standard input \texttt{<query, service>}, the model can spontaneously apply these logics, achieving a transition from ``explicit reliance on reasoning'' to ``implicit autonomous reasoning.''

To achieve this, we propose the \textbf{Contrastive Reasoning Self-Distillation (CRSD)} framework. Its core idea is to model the behavior of the same BERT model under two different input configurations as a teacher-student relationship.
\begin{itemize}[leftmargin=2em,topsep=0pt,partopsep=0pt]
    \item \textbf{Teacher Configuration}: The model receives the reasoning-augmented input \texttt{<query, service, reason>}, and its output \texttt{[CLS]} representation is denoted as $\text{cls}_r$.
    \item \textbf{Student Configuration}: The model receives the standard input \texttt{<query, service>}, and its output \texttt{[CLS]} representation is denoted as $\text{cls}$.
\end{itemize}

Our CRSD framework is depicted in {~\Cref{fig:crsd_framework}. The training process is driven by a combination of three loss functions:

First, we compute standard classification losses for both configurations to ensure the model learns to classify relevance correctly in both modes. We assign a smaller weight $\gamma$ to the teacher's loss to prevent the model from becoming overly reliant on the lengthy reasoning path.
\begin{equation}
    \mathcal{L}_{\text{sce}} = \text{CE}(y_s, y) ,
    \mathcal{L}_{\text{tce}} = \text{CE}(y_r, y)
\end{equation}
where $y_s$ and $y_r$ are the predicted outputs from the student and teacher configurations, respectively, and $y$ is the ground-truth label.
\begin{figure}[h!]
    \centering
    \includegraphics[width=1.0\columnwidth]{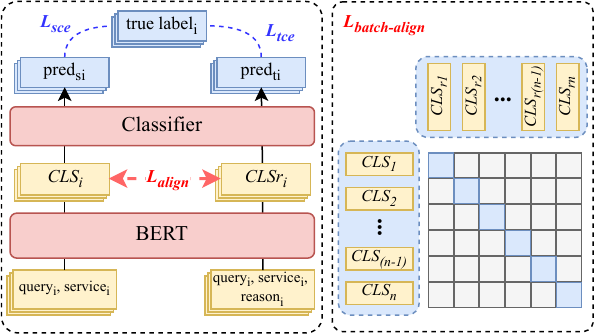}
    \caption{\textbf{Left:} Overview of the proposed CRSD framework. \textbf{Right:} Illustration of the $\mathcal{L}_{\text{align}}$ objective.}
    \label{fig:crsd_framework}
\end{figure}
Second, and most critically, we introduce an in-batch contrastive learning loss to align the semantic representations from the two configurations. Within a single batch, for the $i$-th sample, its student representation $\text{cls}_i$ and teacher representation $\text{cls}_{r_i}$ form a positive pair. The student representation $\text{cls}_i$ and all other teacher representations $\text{cls}_{r_j}$ ($j \neq i$) form negative pairs. By pulling positive pairs closer and pushing negative pairs apart, we compel the model, even with only standard input, to produce a semantic representation similar to the one it would generate if it ``saw'' the reasoning path. 
This loss (InfoNCE) is defined as:
\begin{equation}
    \mathcal{L}_{\text{align}} = -\sum_{i=1}^{N} \log \frac{\exp(\text{sim}(\text{cls}_i, \text{cls}_{r_i})/\tau)}{\sum_{j=1}^{N} \exp(\text{sim}(\text{cls}_i, \text{cls}_{r_j})/\tau)}
\end{equation}
where $\text{sim}(\cdot, \cdot)$ is the cosine similarity and $\tau$ is the temperature hyperparameter.

Finally, the total optimization objective is the weighted sum of these three losses:
\begin{equation}
    \mathcal{L}_{\text{CRSD}} = \mathcal{L}_{\text{sce}} + \gamma \mathcal{L}_{\text{tce}} + \delta \mathcal{L}_{\text{align}}
    \label{eq:crsd_loss}
\end{equation}
where $\gamma$ and $\delta$ are balancing hyperparameters.

We set $\gamma$ and $\delta$ to balance the auxiliary reasoning and alignment tasks with the primary classification goal. This setup ensures the model does not become overly dependent on explicit reasoning paths during training or prioritize representation alignment at the expense of classification accuracy ($\mathcal{L}_{\text{sce}}$). Through this strategy, the reasoning capability is effectively distilled from the teacher configuration and internalized within the student model's parameters for inference.

\section{Experiments}
\subsection{Datasets and Evaluation}

\paragraph{\textbf{Training Datasets.}}
Our training data is composed of two parts:
\begin{itemize}[leftmargin=2em,topsep=0pt,partopsep=0pt]
\item \textbf{LLM Training Data:} The construction of the teacher LLM relies on two types of data: a corpus of \textbf{7 million} domain-specific examples for continued pre-training, and a high-quality, human-annotated dataset totaling over 45,000 instances for supervised fine-tuning and preference optimization. Specifically, the SFT and GRPO stages utilize two independently collected subsets from this annotated dataset—comprising \textbf{34,002} and \textbf{11,634} samples, respectively. These subsets share identical formats and the same source but are mutually exclusive, ensuring that the subsequent preference optimization is performed on novel samples that follow the same distribution as the SFT data, thereby effectively validating its generalizability. 

\item \textbf{Reasoning Distillation Data:} This is used for training the lightweight model. We leverage our fine-tuned LLM to annotate approximately \textbf{2.68 million} <query, service> pairs sampled from Meituan's search logs. For each pair, a three-class relevance label (Irrelevant, Moderate, Relevant) and a corresponding reasoning path are generated. The label distribution is as follows: Irrelevant ($16.2\%$), Moderate ($11.7\%$), and Relevant ($72.1\%$).

\end{itemize}
\paragraph {\textbf{Evaluation Dataset.}} The final performance of all models is assessed on a high-quality, manually annotated test set containing approximately \textbf{12,000} samples. This dataset was cross-validated to ensure label accuracy, and its three-class labeling scheme is consistent with the distillation data.

\paragraph{\textbf{Evaluation Metrics.}}
We employ three standard metrics to evaluate performance: (1) \textbf{Accuracy}, the proportion of correct predictions defined as $\frac{1}{n} \sum_{i=1}^{n} \mathbf{1}(\hat{y}_i = y_i)$, where $\mathbf{1}(\cdot)$ is the indicator function; (2) \textbf{Macro F1}, the unweighted average of F1-scores across all classes; and (3) \textbf{Weighted F1}, the average of class-wise F1-scores weighted by their respective support (sample counts).

\subsection{Implementation Details}

\paragraph{\textbf{Reasoning LLM Training Details.}} We select Qwen3-8B \cite{qwen3} as the base model. During the CPT stage, training is conducted with a batch size of 32 and a learning rate of 1e-5, managed by a cosine scheduler with a warm-up ratio of 0.1. The AdamW optimizer is employed. The model is trained for 2 epochs with a maximum input length of 2,048 tokens, and the final loss converges to approximately 0.6. The hyperparameters for SFT are identical to those used in CPT, with the exception that the batch size is 16 and the number of epochs is set to 3. 
The reward model is trained using the same hyperparameters as the SFT stage. For the subsequent reinforcement learning (RL) stage, we employ the VERL framework~\cite{sheng2025hybridflow}. During training, we sample 16 responses per prompt, using a prompt batch size of 64 (mini-batch size of 32) and the AdamW optimizer with a learning rate of 6e-7. To prevent GPU out-of-memory (OOM) errors, we deploy the reward model using vLLM~\cite{kwon2023efficient} and invoke it via API calls during the RL training process.

\paragraph{\textbf{Reasoning Distillation Details.}} 
For the distillation experiments, we use a 6-layer Chinese BERT as our base student model. We compare our approach against two baselines: (1) a standard data distillation baseline, where BERT is fine-tuned directly on the LLM-generated labels; and (2) an embedding-alignment baseline (Previous Method) as discussed in Section~\ref{sec:baseline_distill}, where the reasoning path $r$ is encoded using the BGE-M3 sentence embedding model \cite{chen2024bge} and aligned with BERT’s \texttt{[CLS]} representation. The alignment loss weight $\mu$ is set to 0.1. Regarding input construction, the standard student configuration follows the \texttt{[CLS] query [SEP] service} format with a maximum length of 64 tokens. For the teacher configuration, we extend the maximum length to 150 by appending the reasoning path (\texttt{[SEP] reason}). All models are trained in a distributed data-parallel environment using the PyTorch \cite{paszke2019pytorch} framework and the AdamW \cite{loshchilov2017decoupled} optimizer. We employ a constant learning rate of 5e-5 and an effective total batch size of 4,096, training for 5,000 steps. For the CRSD method, the balancing hyperparameters for the teacher loss ($\gamma$) and the contrastive loss ($\delta$) are both set to 0.01.

\section{Results}
\subsection{LLM Results and Ablation Study}

We first evaluate the performance of general-purpose LLMs on our domain-specific task to verify the necessity of domain adaptation. As shown in the upper part of Table~\ref{tab:LLM_results}, state-of-the-art models like Qwen3-8B and DeepSeek-R1 (zero-shot and few-shot) exhibit significantly lower performance, with Accuracy hovering between 0.51 and 0.62. These models often struggle with platform-specific terminology and nuanced business logic. This substantial performance gap underscores that general-world knowledge alone is insufficient for high-precision e-commerce search, confirming the vital importance of our proposed domain-adaptive training pipeline.
The bottom part of Table~\ref{tab:LLM_results} presents the results of our reasoning LLM across different stages. It demonstrates a clear and steady performance improvement with each successive training step. Our analysis begins with the SFT Baseline, which represents a model fine-tuned only on human-annotated reasoning data without the benefit of the CPT stage.

The first significant enhancement comes from introducing domain-specific knowledge via Continued Pre-Training. The CPT+SFT model, which builds upon the CPT stage, shows a notable improvement across all metrics, with Macro F1 increasing from 0.6980 to 0.7032. This validates our initial hypothesis: equipping the LLM with foundational knowledge of our platform's services and user behavior through a multi-task CPT phase is crucial before teaching it to perform complex reasoning. Next, we applied GRPO optimization, first using only the final answer's correctness as a reward signal (CPT+SFT+GRPO($R_{label}$)). This led to another substantial performance jump, lifting Macro F1 to 0.7101. This result confirms that reinforcement learning is effective at refining the model's policy. However, as discussed in our methods, relying on a sparse, outcome-only reward risks ``sparse reward'' and ``reward hacking.''

This risk is addressed by our full model, which incorporates the multi-dimensional weighted reward (CPT+SFT+GRPO(Weighted)). This model consistently surpassed all other variants, achieving the highest scores on all metrics and boosting Macro F1 to 0.7174. The significant margin of improvement over the model trained with only the label reward strongly corroborates our central hypothesis: fine-grained calibration of the model's reasoning process, beyond merely focusing on the final answer, leads to a more powerful and reliable teacher model. The success of this three-stage pipeline establishes a robust foundation for the subsequent knowledge distillation stage.

\begin{table}[h!]
\centering
\caption{Offline results for the reasoning LLM and general-purpose LLM baselines. The GRPO variants are applied on top of `CPT+SFT', using either a simple label-based reward (`$R_{label}$') or our full weighted reward (`Weighted'). }
\label{tab:LLM_results}
\begin{tabular}{@{}lccc@{}}
\toprule
\textbf{Method}                  & \textbf{Accuracy} & \textbf{Macro F1} & \textbf{Weight F1} \\ \midrule
Qwen3-8B (Zero-shot)             & 0.5480            & 0.3955            & 0.5748             \\
Qwen3-8B (Few-shot)              & 0.5893            & 0.4375            & 0.6254             \\
DeepSeek-R1 (Zero-shot)          & 0.5140            & 0.4067            & 0.5827             \\
DeepSeek-R1 (Few-shot)           & 0.6197            & 0.4696            & 0.6638             \\ \midrule
SFT Baseline                     & 0.7708            & 0.6980            & 0.7485             \\
CPT+SFT                          & 0.7762            & 0.7032            & 0.7538             \\
+ GRPO ($R_{label}$)             & 0.7812            & 0.7101            & 0.7634             \\
\textbf{+ GRPO (Weighted)}       & \textbf{0.7873}   & \textbf{0.7174}   & \textbf{0.7701}    \\ \bottomrule
\end{tabular}
\vspace{-1em}
\end{table}

\subsection{Distillation Results and Ablation Study}

Having established a powerful teacher model with a Macro F1 of 0.7174 (as shown in Table~\ref{tab:LLM_results}), we evaluate the effectiveness of our distillation framework. The goal is to transfer as much of this reasoning capability as possible to a lightweight 6-layer BERT model suitable for online deployment. While some performance drop is expected when distilling from a massive LLM to a small student, the key is to minimize this gap while significantly outperforming simpler training methods.
We first conducted a detailed ablation study comparing our approach with several baselines. We establish a Baseline using standard data distillation, and a more advanced Baseline Reasoning Distillation method. We then evaluate two variants of our own approach: one using only our contrastive alignment loss (+ $\mathcal{L}_{\text{align}}$) and our complete CRSD (Full) model. The results are summarized in ~\Cref{tab:main_ablation_results}.

\begin{table}[h!]
\centering
\caption{Main results and ablation study of different distillation methods. The best results are highlighted in bold.}
\label{tab:main_ablation_results}
\begin{tabular}{@{}lccc@{}}
\toprule
\textbf{Method}                               & \textbf{Accuracy} & \textbf{Macro F1} & \textbf{Weight F1} \\ \midrule
Baseline                                      & 0.7645          & 0.6945             & 0.7461              \\
Baseline Reasoning Distill.                   & 0.7678          & 0.6959             & 0.7486              \\
CRSD + $\mathcal{L}_{\text{align}}$           & 0.7707          & 0.6993             & 0.7522              \\
\textbf{CRSD Full (ours)}                     & \textbf{0.7761} & \textbf{0.7076}    & \textbf{0.7583}     \\ \bottomrule
\end{tabular}
\end{table}

The results in ~\Cref{tab:main_ablation_results} illustrate the distinct contributions of each component. The \texttt{+ $\mathcal{L}_{\text{align}}$} experiment, which introduces only the in-batch contrastive alignment, already achieves a significant performance boost over the baselines. This is a noteworthy finding, as it demonstrates that forcing the student to mimic its reasoning-augmented state via contrastive learning is an effective strategy for knowledge internalization, even without direct supervision on how the teacher should use the reasoning path. Our final distilled model, CRSD Full, achieves a Macro F1 of 0.7076. \xrz{Notably, McNemar's test results ($p < 10^{-8}$) confirm that the improvement of CRSD over the baseline is statistically significant.} This represents a remarkably successful distillation, as it retains approximately 98.6\% of the teacher LLM's performance in a model that is orders of magnitude smaller and faster.
 
\paragraph{\textbf{Ablation on the Validity of Reasoning Path.}}
Furthermore, to verify that the performance gains are genuinely driven by the \textit{semantic content} of the reasoning paths, rather than being an artifact of the model's structure or longer input sequences, we conducted two additional sanity-check experiments. In the  ``w/o Reasoning Path'' setup, the teacher and student receive identical inputs. In the ``w/ Random Reasoning'' setup, the teacher is fed an irrelevant reasoning path randomly sampled from the entire distillation dataset, simulating the effect of injecting uncorrelated noise into the reasoning process.
The results are shown in Table~\ref{tab:reasoning_path_ablation}.

\begin{table}[h!]
\centering
\caption{Ablation on the importance of reasoning path. }
\label{tab:reasoning_path_ablation}
\begin{tabular}{@{}lccc@{}}
\toprule
\textbf{Method}             & \textbf{Accuracy} & \textbf{Macro F1} & \textbf{Weight F1} \\ \midrule
CRSD Full(Ours)        & 0.7761            & 0.7076            & 0.7583              \\ \midrule
w/o Reasoning Path          & 0.7639            & 0.6947            & 0.7463              \\
w/ Random Reasoning         & 0.7624            & 0.6918            & 0.7440              \\ \bottomrule
\end{tabular}
\end{table}

The results compellingly demonstrate the importance of reasoning. In the \texttt{w/o Reasoning Path} setup, the contrastive loss becomes trivial, and the model's performance regresses to be nearly identical to the standard Baseline. Strikingly, the \texttt{w/ Random Reasoning} experiment shows that providing noisy, irrelevant reasoning is actively detrimental, with performance dropping below the baseline.

\subsection{Online Evaluation}

To validate the practical value of our framework in a real-world business setting, we deployed the CRSD-enhanced lightweight model in Meituan's search advertising system for a rigorous online A/B test. \xrz{Over a two-week period}, 10\% of the live traffic was randomly allocated to the experimental group (our CRSD model), which was compared against a baseline group (the original 6-layer BERT model) comprising 30\% of the traffic. The online experiment demonstrated comprehensive and statistically significant improvements across key business metrics, as summarized in Table~\ref{tab:online_results}.
\begin{table}[h!]
\centering
\caption{Relative improvements of our CRSD model over the baseline in the online A/B test.}
\label{tab:online_results}
\begin{tabular}{@{}lccccc@{}}
\toprule
\textbf{Model} & \textbf{AdCTR} & \textbf{AdCVR} & \textbf{GTV} & \textbf{UVCTR} & \textbf{PVCTR} \\ \midrule
online       & -               & -               & -            & -               & -               \\
Ours    & +0.91\%         & +1.06\%         & +0.40\%      & +0.11\%         & +0.18\%         \\ \bottomrule
\end{tabular}

\end{table}

\xrz{The substantial improvements in Ad CTR and Ad CVR substantiate the model's superior capability in query-ad relevance modeling. By aligning user intent more precisely with platform services, our approach effectively boosted user engagement, which consequently drove a steady increase in the platform's Gross Transaction Volume (GTV). Furthermore, a manual expert review of the top 5 results from 5,521 randomly sampled queries revealed a 30.5 percentage point reduction in the bad case rate. Given these robust results, the model has been successfully deployed in the production environment. }

\section{Conclusion}

In this paper, we proposed a two-stage framework to transfer the reasoning capabilities of Large Language Models (LLMs) to lightweight relevance models for industrial-scale search advertising. Our approach first constructs a domain-adapted reasoning LLM to generate high-quality labels and interpretable reasoning paths. Subsequently, our novel Contrastive Reasoning Self-Distillation (CRSD) method internalizes this reasoning knowledge into a compact BERT model, effectively overcoming the challenges of architectural heterogeneity. Comprehensive offline experiments demonstrate significant improvements in relevance prediction accuracy while maintaining high inference efficiency. Crucially, the offline gains were first validated through rigorous online A/B testing in the Meituan search advertising system, which showed substantial lifts in Ad CTR and Ad CVR, alongside a measurable reduction in the bad case rate. Following these robust results, our framework has been successfully deployed to production environment. This work provides a practical and effective paradigm for integrating advanced LLM reasoning into production-grade systems, bridging the gap between performance, interpretability, and deployability in a real-world e-commerce search advertising context.

\begin{acks}
This research is supported by the Meituan Research Fund, under Grant No. PO250624101698.
\end{acks}

\bibliographystyle{ACM-Reference-Format}
\bibliography{software}

\end{document}